\documentclass[11pt]{elsarticle}
\usepackage{amssymb}
\usepackage{graphicx}
\usepackage{verbatim}  
\usepackage{marvosym}  
\usepackage{amsmath}
\usepackage{amssymb}
\usepackage{euscript}
\usepackage{wasysym}
\usepackage{mathrsfs}
\usepackage{bm}
\usepackage{setspace} 
\usepackage{color}
\usepackage{amsfonts}
\usepackage{geometry}
\usepackage{gensymb}

\journal{Information Sciences}

\begin{document}\begin{spacing}{1.70}
\begin{frontmatter}

\title{\textbf{\begin{Large}QHSL: a quantum hue, saturation, and lightness color model\end{Large}}}

 \author{Fei Yan}
 \author{Nianqiao Li}
 \author{Kaoru Hirota}

\begin{abstract}
Quantum image processing employs quantum computing to capture, manipulate, and recover images in various formats. This requires representations of encoded images using the quantum mechanical composition of any potential computing hardware. In this study, a quantum hue, saturation, and lightness (QHSL) color model is proposed to organize and conceptualize color-assignment attributes using the properties of quantum mechanics (i.e., entanglement and parallelism). The proposed color model is used to define representations of a two-dimensional QHSL image and investigate its data storage, chromatic transformation, and pseudocolor applications. The QHSL representation is introduced for the production of quantum images using triple perceptually relevant components. The efficient use of QHSL images is further explored for applications in the fields of computer vision and image analysis.

\vspace{0.4cm}
\vspace{-0.2cm}
\end{abstract}

\end{frontmatter}

\begin{footnotesize}\begin{spacing}{2.0}

F. Yan and N. Li are with the School of Computer Science and Technology, Changchun University of Science and Technology, Changchun 130022, China (E-mail: yanfei@cust.edu.cn)


K. Hirota is with the Department of Computational Intelligence and Systems Science, Tokyo Institute of Technology, Yokohama 226-8502, Japan

\end{spacing}\end{footnotesize}


\newpage

\section{Introduction}
\label{sec1}
Quantum information science is an interdisciplinary subject spanning physics, mathematics, and computer science \cite{vedral2006introduction}. It involves finding new ways to apply the quantum mechanical effects of nature, particularly superposition and entanglement, to information processing in an attempt to exceed the limits of traditional computing \cite{nielsen2000quantum}. In addition to promoting the mathematical and physical foundations, scientists and engineers have increasingly begun studying cross-disciplinary fields in quantum information processing, such as quantum machine learning \cite{biamonte2017quantum} and quantum neural networks \cite{behrman2000simulations}.

Image processing has become a common task in multiple branches of science and engineering, because of its wide range of military, industrial, and medical applications \cite{gonzalez2009digital}. As such, developing efficient visual information storage and processing algorithms has become an increasingly important and challenging area of research \cite{caraiman2009new}, due primarily to the limitations of classical computer architectures and the complexity of most conventional algorithms \cite{caraiman2013histogram}.

Quantum image processing uses quantum algorithms for storing, processing, and retrieving visual information \cite{yan2020quantum}. The unique properties of quantum information (i.e., computational parallelism) allow quantum technologies to provide improved performance in areas such as computing speed, tamper-proof security, and minimal storage requirements \cite{iliyasu2013towards}.

In 2003, Venegas-Andraca proposed that if an apparatus could detect electromagnetic frequencies and produce a quantum state as output, it could store color in a qubit by translating given frequencies to quantum states \cite{venegas2003storing}. A full image could then be stored in a qubit lattice by updating the indices to specify pixels in the image \cite{venegas2005discrete}. Following this approach, various quantum image representations have been developed, focusing primarily on the color content of the image. Proposals for possible representations have included the flexible representation of quantum images (FRQI) \cite{le2011a}, used to integrate grayscale values and position information in a normalized state. In this process, grayscale information is encoded by a single qubit through the use of angular parameters. The novel enhanced quantum representation (NEQR) \cite{zhang2013neqr} utilizes basis states from a sequence of qubits to denote grayscale pixels and reduce the computational complexity of image reconstruction, thereby providing more accurate information retrieval. The multi-channel representation for quantum images (MCQI) \cite{sun2011a} and the novel quantum representation of color digital images (NCQI) \cite{sang2017anovel} were developed by extending the grayscale information encoded in FRQI and NEQR images to color descriptions. These models use red, green, and blue (RGB) channels to express color information in the image while preserving its normalized state.

The chromatic contents of color quantum images can be encoded by surveying all available quantum image representations \cite{yan2016asurvey}. This can be achieved using an RGB color model since the format can replicate the human eye's sensitivity to red, green and blue. However, the correlation of RGB components with the amount of light incident on an object (and with each other), makes object discrimination difficult \cite{loesdau2014hue}. This limits its application in algorithms such as quantum feature detection \cite{abdel2016new}.

In this study, a quantum hue, saturation, and lightness (QHSL) color model is proposed to organize and conceptualize color-assignment attributes. Hue and saturation are encoded and stored as the two amplitude angles of a qubit. Lightness information is stored as a qubit sequence with a length of \emph{q}, which can be optimized for a given configuration. In addition, we propose the application of this QHSL model to two-dimensional quantum images. This approach is beneficial for depicting color content because the included qubit sequence captures most of the contour detail in the original image. The following sections discuss several basic image processing functions (based on HSL information) and corresponding applications (such as pseudocolor processing).

The remainder of this paper is organized as follows. In Section \ref{sec2}, we introduce the QHSL model and storage retrieval. In Section \ref{sec3}, we discuss QHSL images and several basic image processing operations. In Section \ref{sec4}, we apply pseudocolor processing to QHSL images.

\section{The quantum structure of an HSL color model}
\label{sec2}
\subsection{Basic color models and attributes}\label{sec2-1}
\begin{figure}[h]
  \centering
  \includegraphics[width=15.5cm]{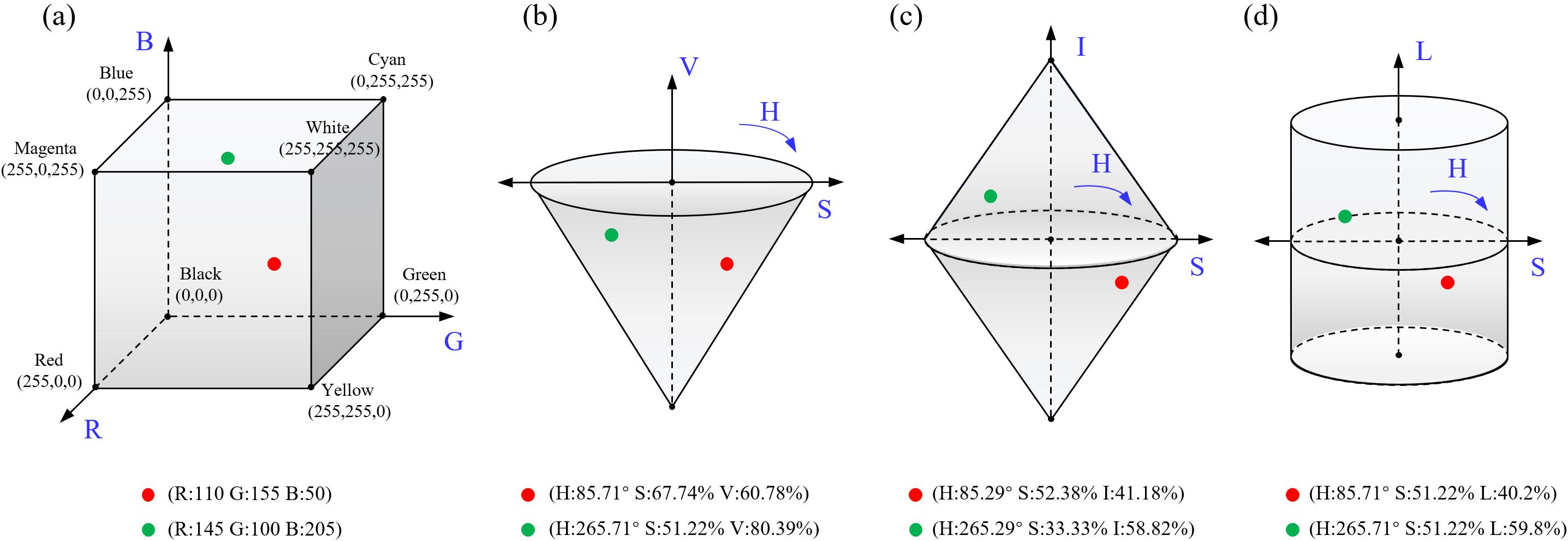}
  \caption{An illustration of (a) RGB, (b) HSV, (c) HSI, and (d) HSL color models.}
  \label{fig1}
\end{figure}

A color model is an abstract mathematical representation of colors as tuples of numbers, typically three or four values indicating specific components \cite{gonzalez2009digital}. When this model is associated with a precise description of how the components are to be interpreted, the resulting set of colors is called a color space. The RGB model, an additive color scheme in which red, green, and blue components are added together to reproduce a broad array of colors, is commonly used in computing \cite{sabine1999standard}. RGB channels can be arranged in a cube, where the tricolors (typically stored as integer numbers in the range 0 to 255) are located on the three axes and the vertices denote eight common colors, as shown in Figure \ref{fig1}(a). Although RGB is intended to replicate the physiology of the human eye, which is sensitive to tricolors, it is incapable of representing human color perception \cite{chen2007identifying}. As such, several alternative models have been proposed that are more closely aligned with natural color attributes \cite{ibraheem2012understanding}. These include the HSV (hue, saturation, value), HSI (hue, saturation, intensity), and HSL (hue, saturation, lightness) models, which can be described as conical, biconical, and cylindrical subsets of the cylindrical coordinate system, respectively. In these models, shown in Figures \ref{fig1}(b)-(d), colors of each hue are arranged in a radial cross-section around a central axis of neutral colors, ranging from black at the bottom to white at the top.

Each of these models can be mathematically converted from RGB \cite{saravanan2016real}. For example, a given color triplet (110, 155, 50) and its complement (145, 100, 205) can be represented as two points within an RGB cube. Corresponding conversion results for the other three models are shown in Figure \ref{fig1}. It is worth noting that while the same name is used for the saturation in each model, they describe substantially different color relationships with the other two components. In addition, significant computational overhead is associated with conversions from HSI and HSV to RGB (and vice versa), since the three color components are strongly correlated. Accordingly, the HSL model is often used in computer vision and image analysis for feature extraction \cite{weeks1995edge} and image segmentation \cite{kalist2015possiblistic}.

\begin{figure}[h]
  \centering
  \includegraphics[width=15cm]{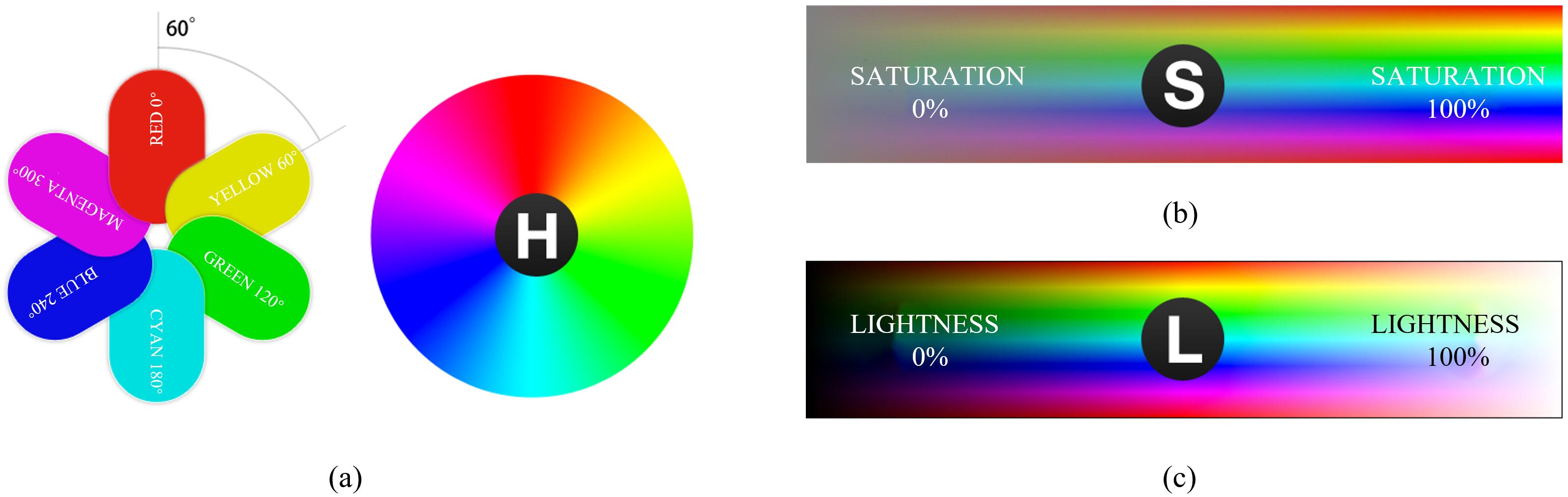}
  \caption{A schematic diagram of the three components in the HSL color model.}
  \label{fig2}
\end{figure}

The angular dimensions of hue vary in the HSL color model, starting with the red primary at 0\degree, passing through the green primary at 120\degree{} and the blue primary at 240\degree, then wrapping back to red at 360\degree{} (as shown in Figure \ref{fig2}). The additive primary and secondary colors (red, yellow, green, cyan, blue, and magenta) and linear mixtures between adjacent pairs, sometimes called pure colors, can be arranged around the outside edge of the cylinder with a saturation of 100\% and a lightness of 50\%. Conversely, the model can be used to represent a variety of grays between black and white when the saturation value is 0\% and the lightness value ranges from 0$\%$ (fully black) to 100$\%$ (fully white) along the central vertical axis.

\subsection{Storage and retrieval in the QHSL model}\label{sec2-2}
\begin{figure}[h]
  \centering
  \includegraphics[width=14cm]{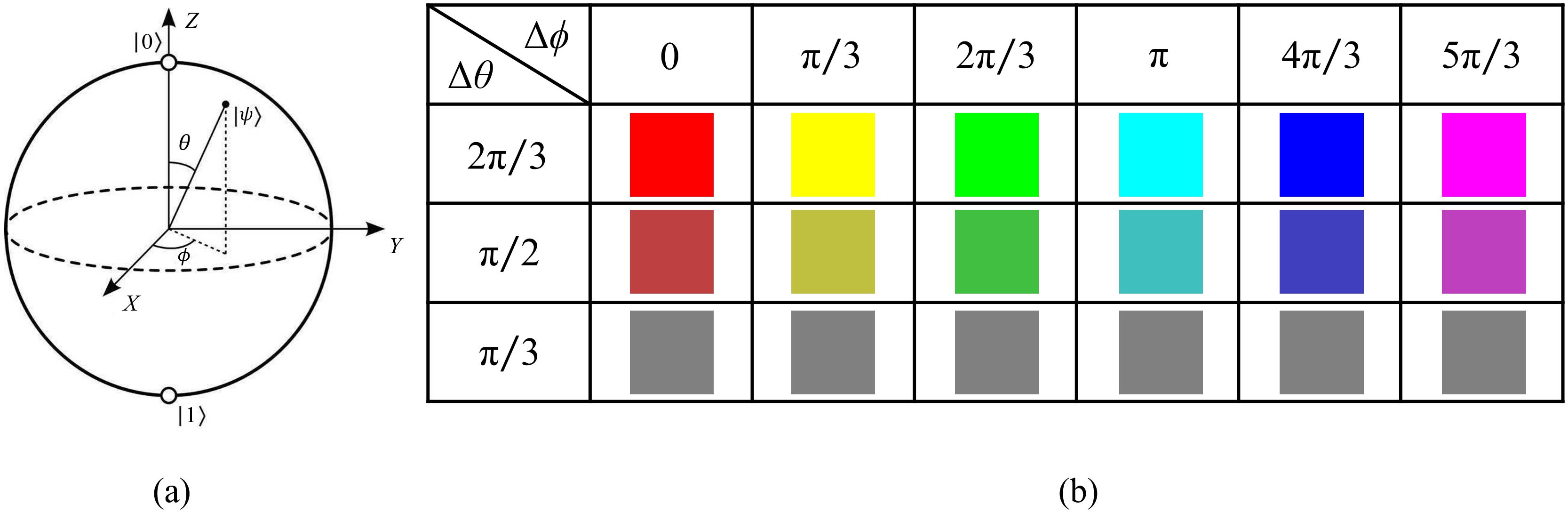}
  \caption{(a) A Bloch sphere composed of the angles $\theta$ and $\phi$. (b) The colors corresponding to different values of $\Delta\theta$ and $\Delta\phi$.}
  \label{fig3}
\end{figure}

In this subsection, a QHSL model defined by chromaticity (further divided into hue and saturation) and lightness is proposed for color representation. In the model, chromaticity is stored in a single qubit and lightness is encoded as a sequence of \emph{q} qubits. The Bloch sphere, shown in Figure {\ref{fig3}}(a), is a geometrical representation of pure state space for a two-level quantum mechanical system (i.e., a qubit) \cite{nielsen2000quantum}. Given an orthonormal basis, any pure state $|\psi\rangle$ of a qubit can be written as a superposition of the basis vectors $|0\rangle$ and $|1\rangle$, where the coefficient or amplitude of each basis vector is a complex number. Quantum mechanics also dictates that the total probability of any system must equal unity: $\langle \psi |\psi \rangle =1$ or equivalently $|||\psi \rangle \,||^{2}=1$. Given this constraint, $|\psi \rangle$ can be expressed using the following representation:
\begin{equation}\label{eq1}
\begin{aligned}
|\psi\rangle&=\cos\frac{\theta}{2}|0\rangle+e^{i\phi}\sin\frac{\theta}{2}|1\rangle\\
&=\cos\frac{\theta}{2}|0\rangle+(\cos\phi+i\sin\phi)\sin\frac{\theta}{2}|1\rangle,
\end{aligned}
\end{equation}
where $\theta\in[0,\pi]$ and $\phi \in [0, 2\pi)$. Since the properties of hue in the HSL model range from $0\degree$ to $360\degree$, it can be represented by a single angle $\phi$. Saturation can be defined by mapping its value (i.e., from 0 to 100\%) to a sub-interval of the angle $\theta$ (i.e., from $\pi/3$ to $2\pi/3$), thereby avoiding possible overflow resulting from the addition and subtraction of saturation \cite{yan2018flexible}. In summary, hue and saturation can be represented in the QHSL model as:
\begin{equation}\label{eq2}
 \begin{aligned}
\phi=\frac{H\times\pi}{180},~\theta=(1+S)\times\frac{\pi}{3},
\end{aligned}
\end{equation}
where the saturation value is set to 0 and 100\% when the angle $\theta$ lies in two bilateral sub-intervals (i.e., $[0, \pi/3]$ and $[2\pi/3, \pi]$), respectively.

As demonstrated in Figures \ref{fig1} and \ref{fig2}, the HSL model attempts to resemble perceptual color models by placing saturated colors in a radial cross-section at a lightness value of 50\%. Hence, a single qubit is capable of storing chromaticity information, as discussed below.

Quantum computers perform calculations by manipulating qubits within a quantum register \cite{nielsen2000quantum}. It is conventionally assumed that an initialized register forms a computational basis state consisting entirely of $|0\rangle$. Changes to this state can be described in the language of quantum computation, using a quantum circuit, in which calculations are performed by a sequence of quantum logic gates (or simply quantum gates). As such, $\theta$ and $\phi$ in Eq. (\ref{eq1}) can be transformed to the desired state using the rotation gates $R_Y(\Delta\theta)$ and $R_Z(\Delta\phi)$ (i.e., rotations around the Y and Z axis of a Bloch sphere by the angles $\Delta\theta$ and $\Delta\phi$, respectively). These operations can be formalized as:
\begin{equation}\label{eq3}
\begin{aligned}
R_Y(\Delta\theta)=
\begin{pmatrix}
\cos\Delta\frac{\theta}{2}&&-\sin\Delta\frac{\theta}{2}\\\sin\Delta\frac{\theta}{2}&&\cos\Delta\frac{\theta}{2}
\end{pmatrix},~
R_Z(\Delta\phi)=
\begin{pmatrix}
1&&0\\0&&e^{i\Delta\phi}
\end{pmatrix}.
\end{aligned}
\end{equation}
Both terms can be transformed simultaneously using:
\begin{equation}\label{eq4}
 \begin{aligned}
R(\Delta\phi,\Delta\theta)=R_Z(\Delta\phi)\times R_Y(\Delta\theta) =
\begin{pmatrix}
\cos\Delta\frac{\theta}{2}&&-\sin\Delta\frac{\theta}{2}\\e^{i\Delta\phi}\sin\Delta\frac{\theta}{2}&&e^{i\Delta\phi}\cos\Delta\frac{\theta}{2}
\end{pmatrix}.
\end{aligned}
\end{equation}

Although $\Delta\theta$ and $\Delta\phi$ are continuous, several typical angles were investigated to identify jointly indicated colors, as shown in Figure \ref{fig3}(b). For example, when $\Delta\theta=2\pi/3$ and $\Delta\phi=2\pi/3$, Eq. (\ref{eq2}) yields H = $120\degree$ and S = 100\%. This corresponds to an indicated color of pure green, with a lightness of 50\% by default. As such, the use of quantum rotation gates involving the angles $\Delta\phi$ and $\Delta\theta$ allows chromaticity information in the QHSL model to be stored within a single qubit.

\begin{figure}[h]
  \centering
  \includegraphics[width=16cm]{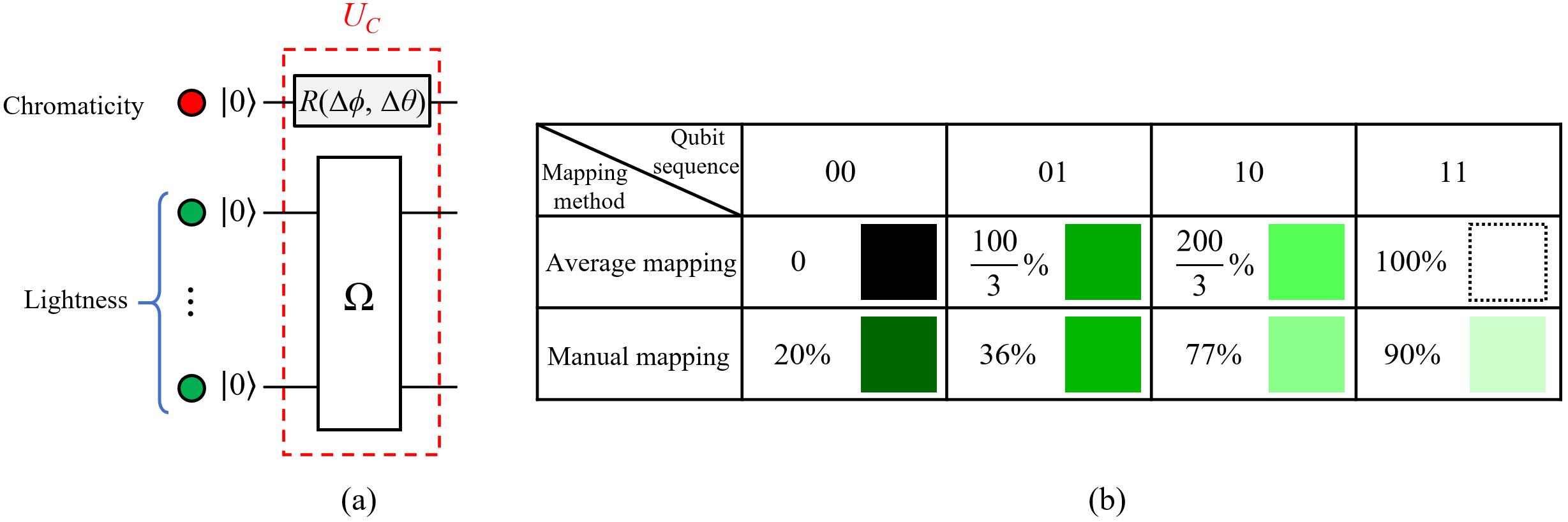}
  \caption{(a) A preparation circuit diagram for the QHSL model. (b) Colors corresponding to different lightness value mapping methods.}
  \label{fig4}
\end{figure}

The lightness component is independent of chromaticity and is more robust for describing color details. As a result, storing lightness values as a qubit sequence makes the model more adaptable. Similar to chromaticity, qubits in the lightness sequence are initialized as $|0\rangle$, symbolically expressed as $|0\rangle^{\otimes q}$ (denoting the repeated tensor product of a basis state $\vert 0\rangle$ \emph{q} times). Lightness information can then be transformed from its initialized value $|0\rangle^{\otimes q}$ to a desired quantum color state using the set operation $\Omega$ as follows:
\begin{equation}\label{eq5}
 \begin{aligned}
\Omega |0\rangle^{\otimes q}&=\bigotimes_{i=0}^{q-1}(\Omega^{i}|0\rangle)=\bigotimes_{i=0}^{q-1}|0 \oplus L^i\rangle=\bigotimes_{i=0}^{q-1}|L^i\rangle,
\end{aligned}
\end{equation}
where $\oplus$ is the XOR operation applied to each qubit in the initial quantum sequence ($\vert 0\rangle^{\otimes q}$) and the corresponding target sequence ($\bigotimes_{i=0}^{q-1}|L^i\rangle$). The QHSL model can thus be initialized using the color-setting operation $U_C$, defined as:
\begin{equation}\label{eq6}
\begin{aligned}
U_C=R(\Delta\phi,\Delta\theta)\otimes\Omega,
\end{aligned}
\end{equation}
as shown in the preparation circuit diagram of Figure \ref{fig4}(a). The complete QHSL model can therefore be expressed as:
\begin{equation}\label{eq7}
\begin{aligned}
|HSL\rangle &= \vert HS\rangle\otimes\vert L\rangle=(\cos\frac{\theta}{2}|0\rangle+e^{i\phi}\sin\frac{\theta}{2}|1\rangle)\bigotimes_{i=0}^{q-1}|L^i\rangle,
\end{aligned}
\end{equation}
where $\theta\in [0,\pi]$ indicates saturation, $\phi\in[0,2\pi)$ indicates hue, and $L^{0}, L^{1}, \cdots, L^{q-1}\in[0,2^q-1]$, where $L^i\in\{0, 1\}$ denotes lightness. Lightness values can be mapped from 0 to 100\% using a quantum sequence consisting of \emph{q} qubits. Two different approaches were used for this mapping, in which \emph{q} qubits typically store $2^q$ values. One technique is an average mapping and the other is a manual mapping, in which a mapping table is constructed artificially in advance. The pure green color in Figure \ref{fig3}(b) can be used as an example to illustrate the influence of lightness in the QHSL model. If we let \emph{q} = 2, the qubit sequence can map 4 lightness values. The corresponding color change is shown in Figure \ref{fig4}(b) for each value.

Simplicity and efficiency of storage and retrieval are essential requirements for representing color information in a quantum model \cite{iliyasu2012watermarking}. Storage was discussed in the previous section and the following section describes QHSL retrieval. Quantum measurements are a unique tool used to recover classical information from a quantum system \cite{li2018colorimage}. As with QHSL storage, measurement-based retrieval can be divided into two steps.

First, $|HS\rangle=\cos\frac{\theta}{2}|0\rangle+e^{i\phi}\sin\frac{\theta}{2}|1\rangle$ can be measured using the computational basis $\{|0\rangle, |1\rangle\}$, such that the probability of measuring 0 is $P_0=|\langle0|HS\rangle|^2=\cos^2\frac{\theta}{2}$ and the probability of measuring 1 is $P_1=|\langle1|HS\rangle|^2=\sin^2\frac{\theta}{2}$. Therefore,
\begin{equation}\label{eq8}
\begin{aligned}
P_0-P_1=\cos^2\frac{\theta}{2}-\sin^2\frac{\theta}{2}=\cos \theta=K,\\
\end{aligned}
\end{equation}
yielding the angle $\theta=\arccos(K)$, where $\theta\in[0,\pi]$. In quantum computation, all transforms are unitary in nature and are typically described using unitary matrices \cite{nielsen2000quantum}. A matrix is said to be unitary if its Hermitian conjugate or its adjoint is equivalent to its inverse. To compute the angle $\phi$, we consider the following two unitary matrices:
\begin{equation}\label{eq9}
\begin{aligned}
&U_1=\frac{1}{\sqrt{2}}\begin{pmatrix}
1&&1\\-1&&1
\end{pmatrix},~U_2=\frac{1}{\sqrt{2}}\begin{pmatrix}
1&&-i\\-i&&1
\end{pmatrix}.\\
\end{aligned}
\end{equation}
By operating on $|HS\rangle$ with $U_1$, we achieve the following result:
\begin{equation}\label{eq10}
|HS'\rangle=U_1|HS\rangle=\frac{1}{\sqrt{2}}\begin{pmatrix}
\cos\frac{\theta}{2}+e^{i\phi}\sin\frac{\theta}{2}\\-\cos\frac{\theta}{2}+e^{i\phi}\sin\frac{\theta}{2}
\end{pmatrix}.\\
\end{equation}
Measurements on $\vert HS'\rangle$ produce outputs of 0 and 1 with probabilities of $P_0'=|\langle0|HS'\rangle|^2=\frac{1}{2}(1+\cos\phi \sin \theta)$ and $P_1'=|\langle1|HS'\rangle|^2=\frac{1}{2}(1-\cos\phi \sin \theta)$. This gives:
\begin{equation}\label{eq11}
 \begin{aligned}
P_0'-P_1'=\cos\phi \sin\theta=V.\\
 \end{aligned}
\end{equation}
Similarly, operating on $\vert HS\rangle$ with $U_2$ produces the quantum state $|HS''\rangle$:
\begin{equation}\label{eq12}
 \begin{aligned}
|HS''\rangle=&U_2|HS\rangle=\frac{1}{\sqrt{2}}\begin{pmatrix}
\cos\frac{\theta}{2}+e^{i(\phi-\frac{\pi}{2})}\sin\frac{\theta}{2}\\e^{i(-\frac{\pi}{2})}\cos\frac{\theta}{2}+e^{i\phi}\sin\frac{\theta}{2}
\end{pmatrix},\\
 \end{aligned}
\end{equation}
and the probability distributions:
\begin{equation}\label{eq13}
 \begin{aligned}
P_0''-P_1''=\sin\phi \sin \theta=W.\\
 \end{aligned}
\end{equation}

The term $\phi=\arctan(W/V)\in[-\pi/2, \pi/2]$ can be calculated from the resulting values of \emph{V} and \emph{W}. The range of $\phi$ can be restored to its proper value of $[0, 2\pi)$, as the hue information requires, using the following operations:
\begin{equation}\label{eq14}
 \begin{aligned}
&\phi=\left\{
\begin{aligned}
 &\arctan(\frac{W}{V})  &&V\ge0, W\ge0\\
 &2\pi+\arctan(\frac{W}{V}) &&V\ge0, W \textless 0\\
 &\pi+\arctan(\frac{W}{V}) &&V\textless0\\
 \end{aligned}
\right..\\
 \end{aligned}
\end{equation}
These results for $\theta$ and $\phi$ can be used in Eq. (\ref{eq2}) to determine the values of hue and saturation in the QHSL model. In addition, the projection matrix $M = \sum_{m=0}^{2^q-1} m|m\rangle\langle m|$ is required for measurement of the qubit sequence $|L\rangle$. This process can be described as follows:
\begin{equation}\label{eq15}
\begin{aligned}
\langle L|M|L\rangle &= \sum_{m=0}^{2^q-1}m\bigotimes_{i=0}^{q-1}\langle L^i||m\rangle\langle m|\bigotimes_{i=0}^{q-1}|L^i\rangle=L^0L^1\cdots L^{q-1}.
\end{aligned}
\end{equation}

Measurements cause the superposition state in quantum systems to collapse \cite{nielsen2000quantum}. Thus, in order to obtain a reasonable estimation of hue and saturation information in the QHSL model, the quantum state $\vert HS\rangle$ must be established multiple times. In addition, multiple quantum measurements of these states produce a series of readouts that form a histogram, implicitly representing the corresponding probability distributions. Extracting and analyzing these distributions provides an estimate of hue and saturation values in the model, which is composed of a single qubit and a qubit sequence used to store color information. When combined with other data (such as pixel position), the QHSL model could be used directly for applications in quantum image processing.

\section{Applications of the QHSL model to two-dimensional images}
\label{sec3}
\subsection{Representation of a QHSL image}
\label{sec3-1}
Quantum image representation uses a mathematical model to manipulate pixels \cite{yan2016asurvey}. Like classical images, quantum images consist of both color and position information. This can be expressed as:
\begin{equation}\label{eq16}
 \begin{aligned}
|I\rangle=\frac{1}{\sqrt{N}}\sum_{P=0}^{N-1}|C_P\rangle\otimes|P\rangle,
\end{aligned}
\end{equation}
where $|C_P\rangle$ represents the color and $|P\rangle$ represents the corresponding position information. Image data can be integrated into a normalized quantum state using the tensor product $\otimes$. In a two-dimensional (2D) Cartesian coordinate system:
\begin{equation}\label{eq17}
 \begin{aligned}
|P\rangle=|Y\rangle|X\rangle=|Y_{n-1}Y_{n-2}\cdots Y_0\rangle|X_{n-1}X_{n-2}\cdots X_0\rangle,
\end{aligned}
\end{equation}
where $X_j, Y_j \in \{0,1\}$. In this study, $|C_P\rangle$ denotes the QHSL model developed in the previous section. A 2D QHSL image can then be expressed as:
 \begin{equation}\label{eq18}
 \begin{aligned}
|I_{HSL}\rangle&={1\over 2^{n}}\sum_{Y=0}^{2^n-1}\sum_{X=0}^{2^n-1}\vert HSL(Y, X)\rangle\otimes\vert YX\rangle\\
&={1\over 2^{n}}\sum_{Y=0}^{2^n-1}\sum_{X=0}^{2^n-1}(\cos\frac{\theta_{YX}}{2}|0\rangle+e^{i\phi_{YX}}\sin\frac{\theta_{YX}}{2}|1\rangle)\bigotimes_{i=0}^{q-1}|L_{YX}^i\rangle|YX\rangle.
\end{aligned}
\end{equation}

Transforming a quantum computer from its initialized configuration to a desired state is the first step in quantum image representation and processing \cite{li2019quantumimplementation}. The following sections discuss core requirements for developing a QHSL image. The initialized quantum system is assumed to include $2n+q+1$ basis states $\vert 0\rangle$ (denoted by $\vert 0\rangle^{\otimes {2n+q+1}}$), where \emph{q} qubits are used to encode lightness information and $2n$ qubits are used to encode position information in a $2^n\times 2^n$ QHSL image. The remaining qubit is used to encode hue and saturation data. The preparation of a QHSL image primarily consists of two steps.

\textbf{Step 1:} The transformation $U_P=I^{\otimes q+1}\otimes H^{\otimes 2n}$ is applied to the initialized sate $|\psi\rangle_0=|0\rangle^{\otimes 2n+q+1}$ to produce an intermediate state $|\psi\rangle_1$ as follows:
\begin{equation}\label{eq19}
\begin{aligned}
|\psi\rangle_1=U_P|\psi\rangle_0=\frac{1}{2^n}\sum_{Y=0}^{2^n-1}\sum_{X=0}^{2^n-1}|0\rangle^{\otimes q+1}|YX\rangle.
\end{aligned}
\end{equation}
Two unitary matrices, the 2D identity matrix  $I=\begin{pmatrix}1&&0\\0&&1\end{pmatrix}$ and a Hadamard matrix $H=\frac{1}{\sqrt{2}}\begin{pmatrix}1&&1\\1&&-1\end{pmatrix}$, are included in this process. Once position information has been initialized, the intermediate state $|\psi\rangle_1$ can be considered the superposition of all pixels in an empty quantum image with all color values set to $|0\rangle$.

\textbf{Step 2:} The color-setting operation $U_C$ defined in Eq. (\ref{eq6}) is used to generate color values for each pixel. The quantum sub-operation $U_{YX}$ applied to initialize color information for a pixel at position (Y, X) can be expressed as:
\begin{equation}\label{eq20}
\begin{aligned}
U_{YX}=\bigg(I\otimes\sum_{ji=0, ji\not=YX}^{2^{2n}-1}|ji\rangle\langle ji|\bigg)+U_C\otimes|YX\rangle\langle YX|.
\end{aligned}
\end{equation}

Applying $U_{YX}$ to this intermediate state $|\psi\rangle_1$ gives:
\begin{equation}\label{eq21}
 \begin{aligned}
U_{YX}|\psi\rangle_1&=U_{YX}\bigg(\frac{1}{2^n}\sum_{ji=0}^{2^{2n}-1}|0\rangle^{\otimes q+1}|ji\rangle\bigg)
\\&=\frac{1}{2^n}\bigg(\sum_{ji=0, ji\not=YX}^{2^{2n}-1}|0\rangle^{\otimes q+1}|ji\rangle\\&+R(\Delta\phi_{YX},\Delta\theta_{YX})|0\rangle\otimes\Omega_{YX}|0\rangle^{\otimes q}|YX\rangle\bigg)
\\&=\frac{1}{2^n}\bigg(\sum_{ji=0, ji\not=YX}^{2^{2n}-1}|0\rangle^{\otimes q+1}|ji\rangle\\&+\bigg(\cos\frac{\theta_{YX}}{2}|0\rangle+e^{i\phi_{YX}}\sin\frac{\theta_{YX}}{2}|1\rangle\bigg)\otimes|L_{YX}\rangle|YX\rangle\bigg).
\end{aligned}
\end{equation}

As $U_{YX}$ can only be applied to one pixel at a time, $2^{2n}$ sub-operations defined as:
\begin{equation}\label{eq22}
 \begin{aligned}
U_{I}=\prod_{YX=0}^{2^{2n}-1}U_{YX},
\end{aligned}
\end{equation}
are required to process a $2^n\times 2^n$ quantum image. The operation $U_I$ can be used to transform the quantum state $|\psi\rangle_1$ into the final QHSL image state $|\psi\rangle_2$ as follows:
\begin{equation}\label{eq23}
 \begin{aligned}
|\psi\rangle_2&=U_{I}\big(|\psi\rangle_1\big)
\\&=\frac{1}{2^n}\bigg(\sum_{YX=0}^{2^{2n}-1}R(\Delta\phi_{YX},\Delta\theta_{YX})|0\rangle\bigotimes\Omega_{YX}|0\rangle^{\bigotimes q}|YX\rangle\bigg)\\&=\frac{1}{2^n}\bigg(\sum_{YX=0}^{2^{2n}-1}|HSL(Y,X)\rangle|YX\rangle\bigg).
\end{aligned}
\end{equation}

\begin{figure}[h]
  \centering
  \includegraphics[width=15cm]{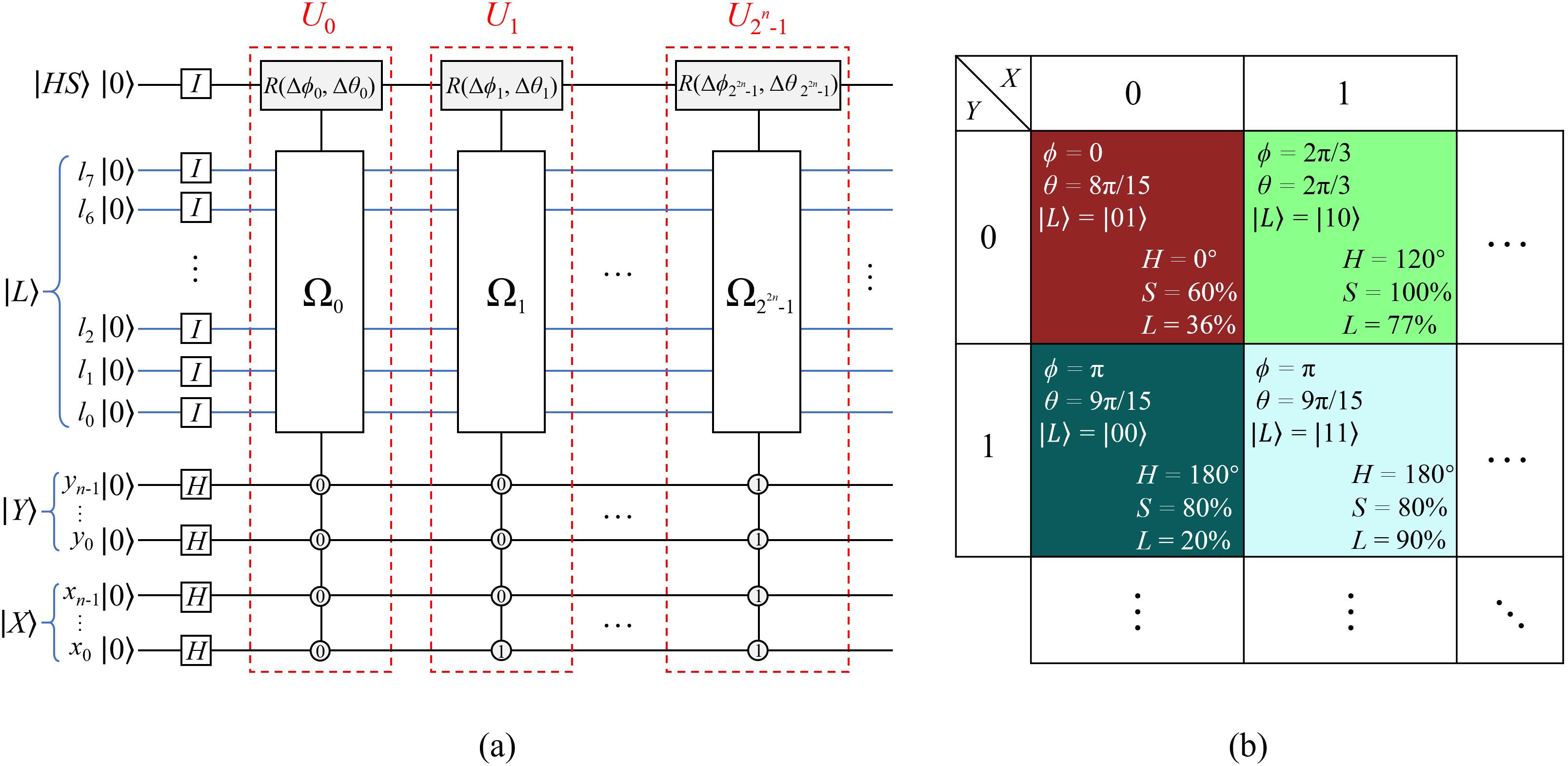}
  \caption{(a) A quantum circuit for QHSL image preparation. (b) An example of a $2\times 2$ QHSL image.}
  \label{fig5}
\end{figure}

A quantum circuit for the preparation of a QHSL image is shown in Figure \ref{fig5}(a), where the red dashed rectangles denote the operations $U_{0}, U_{1}, \cdots, U_{2^{2n}-1}$ used to initialize each pixel in the image. Hue, saturation, and lightness information can then be represented by a single qubit, as discussed in Section \ref{sec2-2} (see Figures \ref{fig3} and \ref{fig4}). Here we provide an example of a QHSL image, in which the color information in a $2\times 2$ sub-area is initialized as shown in Figure \ref{fig5}(b).

In combination with the color data retrieved from the QHSL model (see Section \ref{sec2-2}), the quantum measurement $\Gamma$ can be applied to a position qubit sequence to extract corresponding position information as follows:
\begin{equation}\label{eq24}
\begin{aligned}
\Gamma=\sum_{YX=0}^{2^{2n}-1}I^{\bigotimes q+1}\bigotimes|YX\rangle\langle YX|.
\end{aligned}
\end{equation}
These measurement operations can be used to extract color information for all pixels in the QHSL image, thereby reconstructing the classical image.

\begin{figure}[h]
  \centering
  \includegraphics[width=15cm]{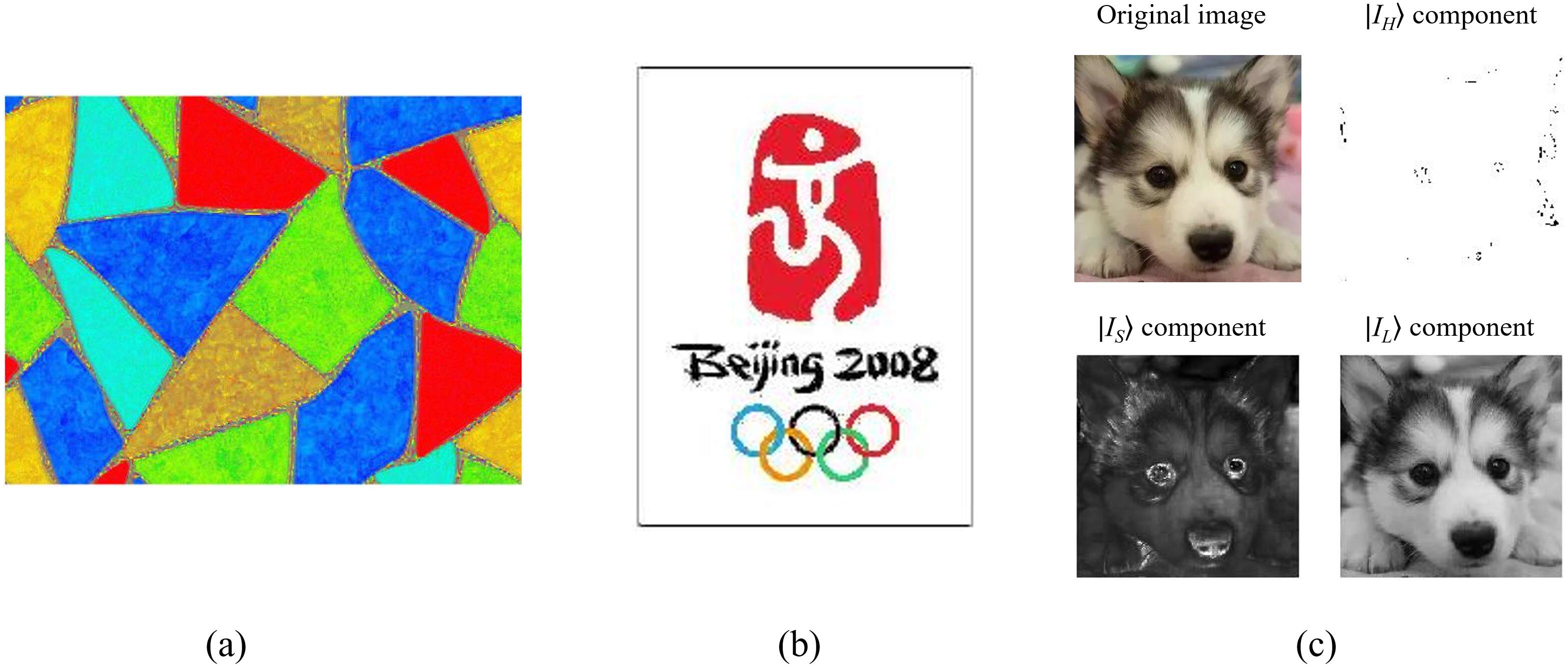}
  \caption{(a) A tiled image that includes only one lightness value. (b) The Beijing Olympic emblem reproduced with only three lightness values. (c) The three channels of a QHSL image depicting a pet dog.}
  \label{fig6}
\end{figure}

As mentioned previously, the number of qubits \emph{q} used to denote lightness information $|L\rangle$ can vary for each individual application. For example, we assume Figure \ref{fig6}(a) represents a QHSL image containing only color information with a default lightness of 50\%. The corresponding color requires only a single chromaticity qubit (i.e., hue and saturation), without any additional qubits to indicate lightness (\emph{q}=0). In addition, Figure \ref{fig6}(b) shows a QHSL image with pixel lightness values of 0\% (black), 50\%, and 100\% (white). This type of representation only requires two qubits.

An optimal value of \emph{q} can be chosen to better represent image details and enhance color. For example, in Figure \ref{fig6}(c), the original QHSL image can be decomposed into hue, saturation, and lightness. From this, it is evident that the $\vert I_H\rangle$ and $\vert I_S\rangle$ component images only describe color properties, while $\vert I_L\rangle$ fully describes contour information, as shown in Figure \ref{fig6}(c). By comparing the original image with its lightness component, it is observed that $\vert I_L\rangle$ is similar to the grayscale version of the original color image.

Grayscale pixels use a single value to represent image brightness at a specific location, ranging from 0 (black) to 1 (white). Pixel depth allows for 256 different intensities, typically stored using 8 bits per sampled pixel. Similarly, the proposed QHSL model uses 8 qubits to divide lightness information into 256 intervals, which allows for simple processing of grayscale quantum images.

Quantum image processing requires the development of mathematical operators (e.g., the quantum Fourier transform \cite{weinstein2001implementation}) to convert an input image into a different representation \cite{li2019quantumvision}. In this process, the expected results and the associated computational circuits must be understood prior to processing. The following two subsections introduce possible operations for QHSL images.

\subsection{Global QHSL image operations}\label{sec3-2}
Global operations are applied to all pixels in a QHSL image. The process of altering color information defined by hue, saturation, lightness, or any combination of these requires a global operation. This process is discussed in more detail below.

(1) Transforming the hue component in QHSL images

\begin{figure}[h]
  \centering
  \includegraphics[width=15cm]{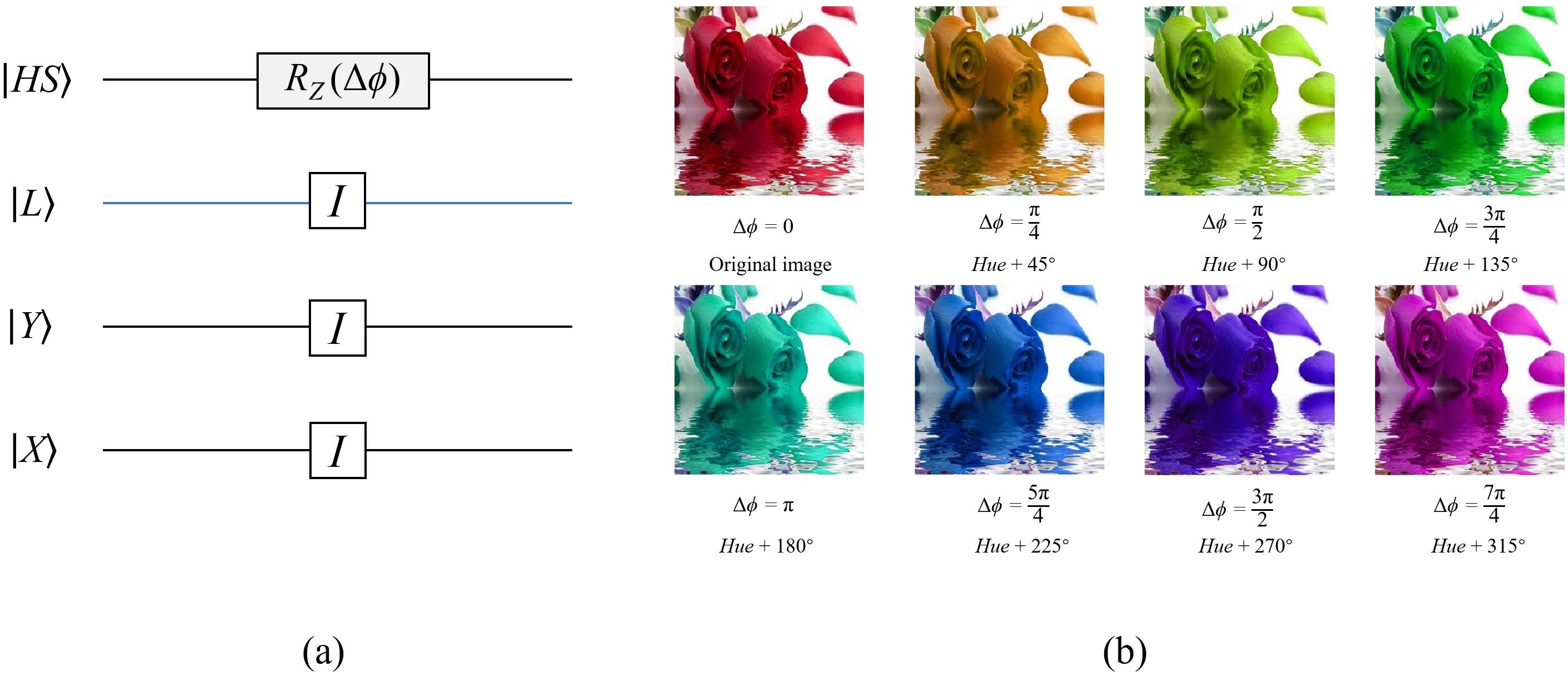}
  \caption{(a) A quantum circuit diagram that modifies only the hue component. (b) The effects of varying $\Delta\phi$ on a sample QHSL image.}
  \label{fig7}
\end{figure}

Modifying hue requires the use of a rotation matrix $R_Z(\Delta\phi)$, defined in Eq. (\ref{eq3}). A quantum circuit diagram that produces changes in only the hue component of QHSL images is shown in Figure \ref{fig7}(a). Figure \ref{fig7}(b) shows the effects of varying $\Delta\phi$ on a sample image, in which the rose color is observed to change sequentially from red to magenta. The color formed when $\Delta\phi=\pi$ is referred to as the complementary color of the original image.

(2) Transforming the saturation component

\begin{figure}[h]
  \centering
  \includegraphics[width=15.5cm]{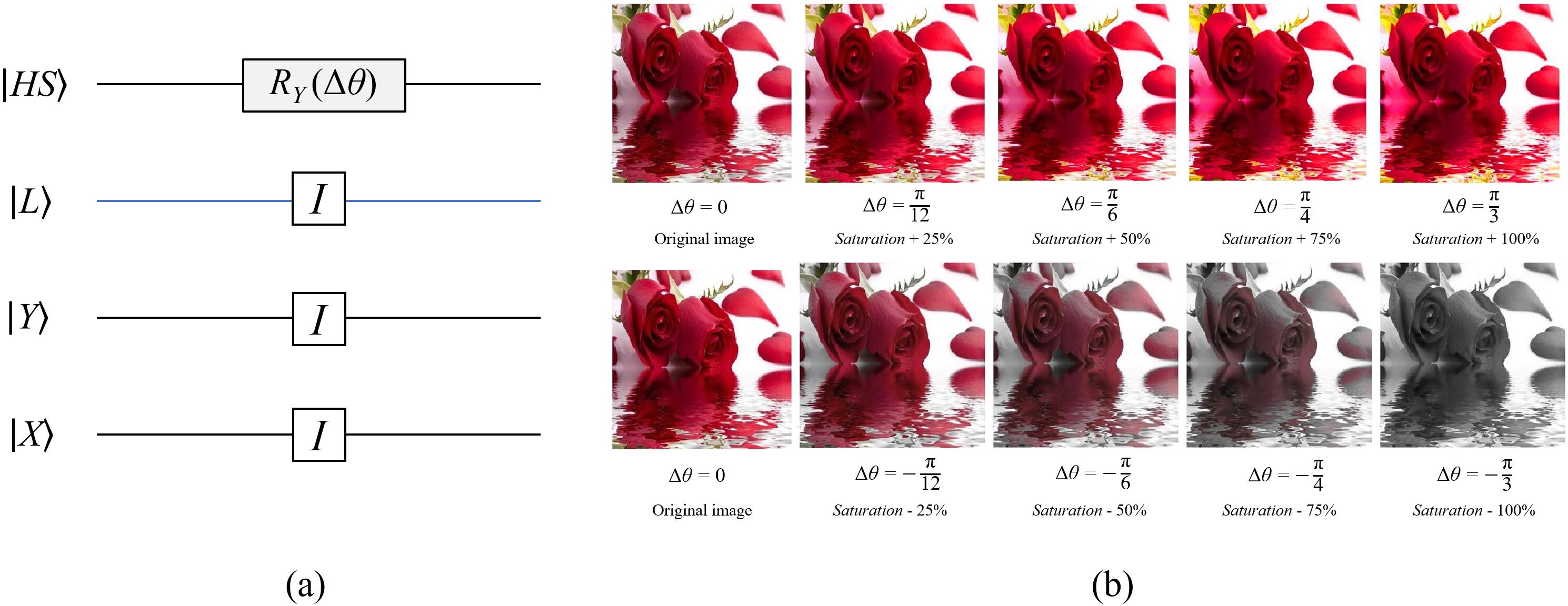}
  \caption{(a) A quantum circuit diagram that changes only the saturation component. (b) The effects of varying $\Delta\theta$ on a sample QHSL image.}
  \label{fig8}
\end{figure}

Modifying saturation requires the use of a rotation matrix $R_Y(\Delta\theta)$, defined in Eq. (\ref{eq3}). A quantum circuit diagram that changes only the saturation component of a QHSL image is shown in Figure \ref{fig8}(a). Figure \ref{fig8}(b) shows the effects of varying $\Delta\theta$ for a sample image. As $\Delta\theta$ increases, the color of the rose becomes increasingly saturated until all pixels reach a value of $100\%$ at $\Delta\theta =\pi/3$. Conversely, the color dulls for negative values until the saturation reaches 0 at $\Delta\theta =-\pi/3$. Addition and subtraction operations can produce a value for $\Delta\theta$ outside the defined interval of [$\pi/3, 2\pi/3$]. When this occurs, the saturation is truncated at either $0\%$ (when the lower limit is breached) or $100\%$ (when the upper limit is breached), as discussed in Section \ref{sec2-2}.

(3) Transforming the lightness component

\begin{figure}[h]
 \centering
 \includegraphics[width=12.5cm]{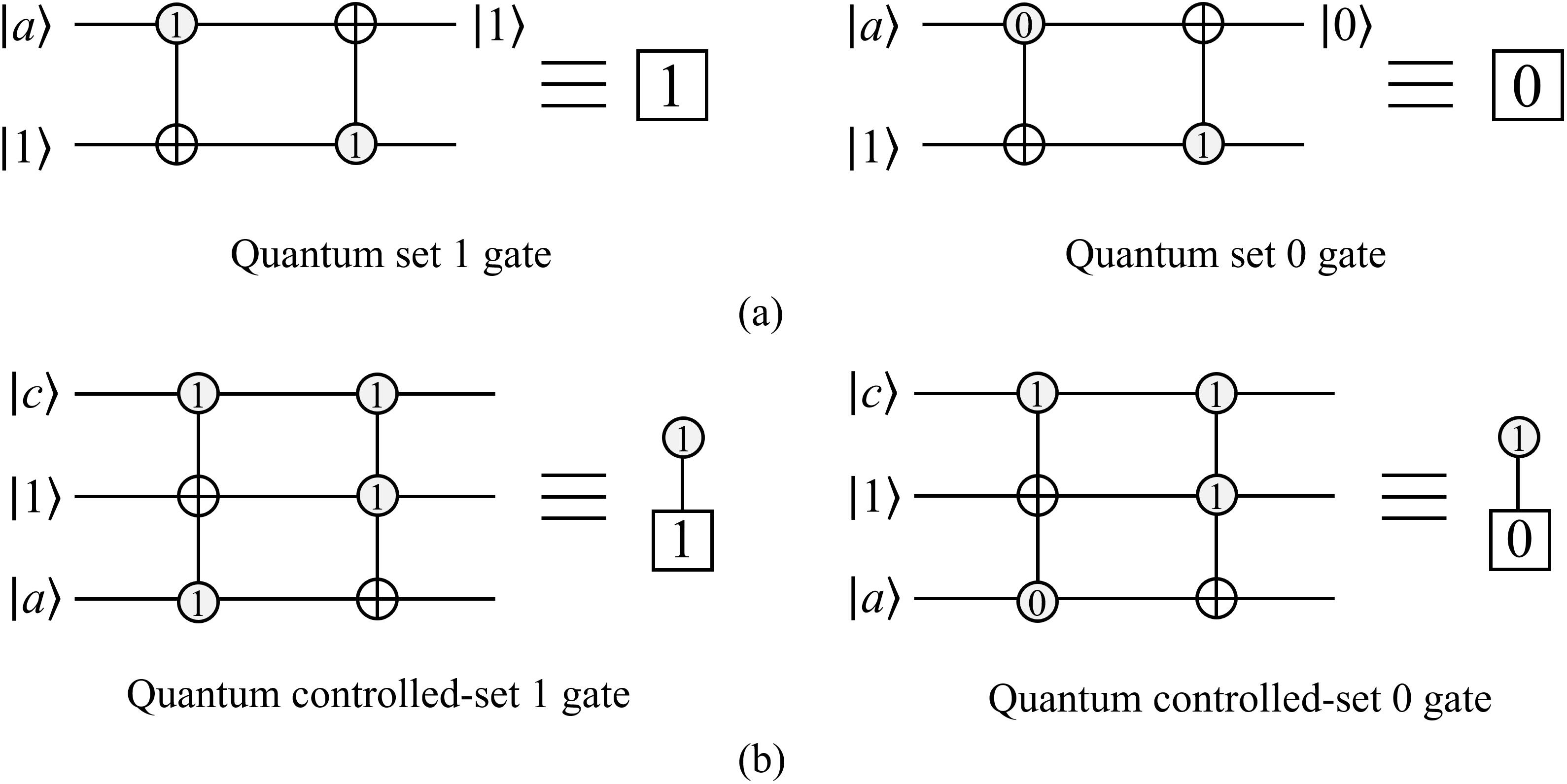}
 \caption{(a) Quantum set 1 and 0 gates. (b) Quantum controlled-set 1 and 0 gates.}
 \label{fig9}
\end{figure}

Lightness information is encoded by a sequence of qubits and its modification requires operating on multiple qubits, setting them to either $|1\rangle$ or $|0\rangle$. For example, the circuit operation shown in Figure \ref{fig9}(a) can be used to set an input qubit $\vert a\rangle$ to either state. Figure \ref{fig9}(b) shows a circuit used to perform such operations with a control condition $\vert c\rangle$.

\begin{figure}[h]
 \centering
 \includegraphics[width=16.5cm]{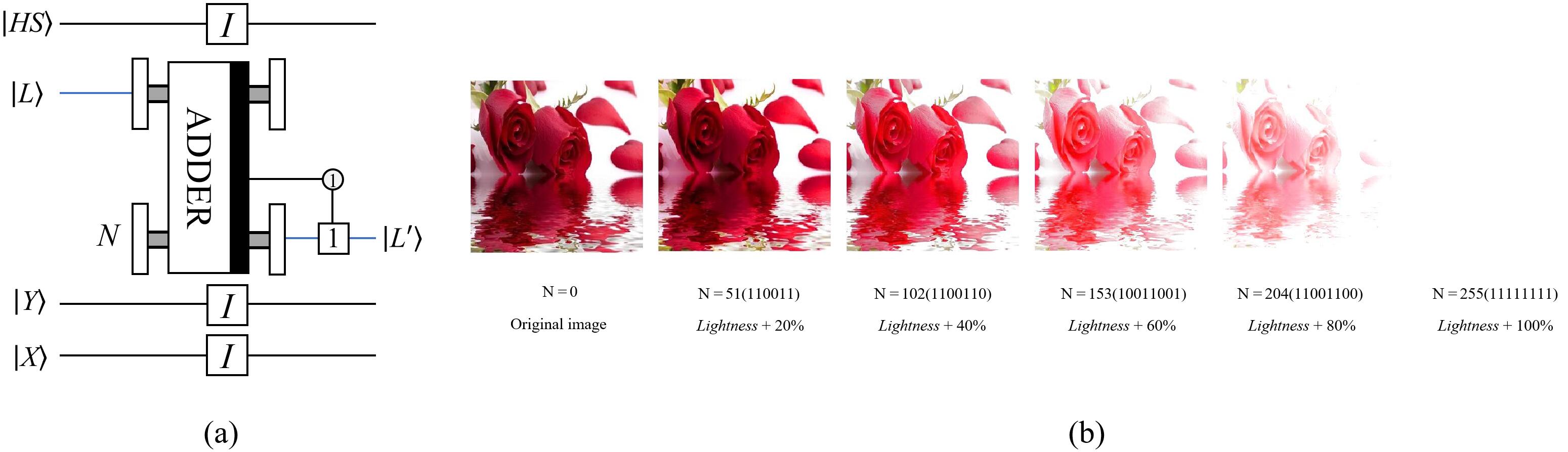}
 \caption{(a) A quantum circuit diagram that changes only the lightness component. (b) The effects of varying lightness on a sample QHSL image.}
 \label{fig10}
\end{figure}

Figure \ref{fig10}(a) shows a quantum circuit diagram used for the addition of lightness components to QHSL images with an integer value. This addition operation is conducted using a quantum adder module \cite{vlatko1996quantum}, which consists of 2\emph{n} sum and 2\emph{n}-1 carry sub-modules for bitwise summation and carries. A maximum quantum adder carry bit of 1 indicates that the addition of two inputs exceeds the maximum value of an 8-qubit binary number (i.e., 11111111 or 255). As such, the quantum controlled-set 1 gate will be used to set the value of all qubits representing the lightness component to 1. In other words, the lightness value of the pixel is directly set to 255 (100\%). The effects of changing lightness values are shown for a sample QHSL image in Figure \ref{fig10}(b). It is evident from the figure that higher lightness produces brighter colors that are closer to white.

As quantum gates are reversible, subtraction could be implemented using a network of quantum adders. An integer value can be subtracted from the lightness component of a QHSL image using the corresponding quantum controlled-set 0 gate. In this process, the color of an image becomes increasingly darker, which will not be illustrated here.

The operations discussed above apply to only one of three image components. As mentioned previously, some combinations of operations on $\vert H\rangle$, $\vert S\rangle$, and $\vert L\rangle$ produce unique results. For example, the combination of quantum NOT gates and a rotation matrix $R_Z(\Delta\phi=\pi)$ applied to lightness and hue values constitutes an inverse color operation.

\subsection{Local QHSL image operations}\label{sec3-3}
Transformations can be confined to smaller sub-areas or a specified lightness range within an image, by imposing additional constraints on the position or lightness qubit sequences. These operations are defined in this subsection using simple control conditions and quantum comparators.

(1) Local operations with simple control conditions

\begin{figure}[h]
  \centering
  \includegraphics[width=17cm]{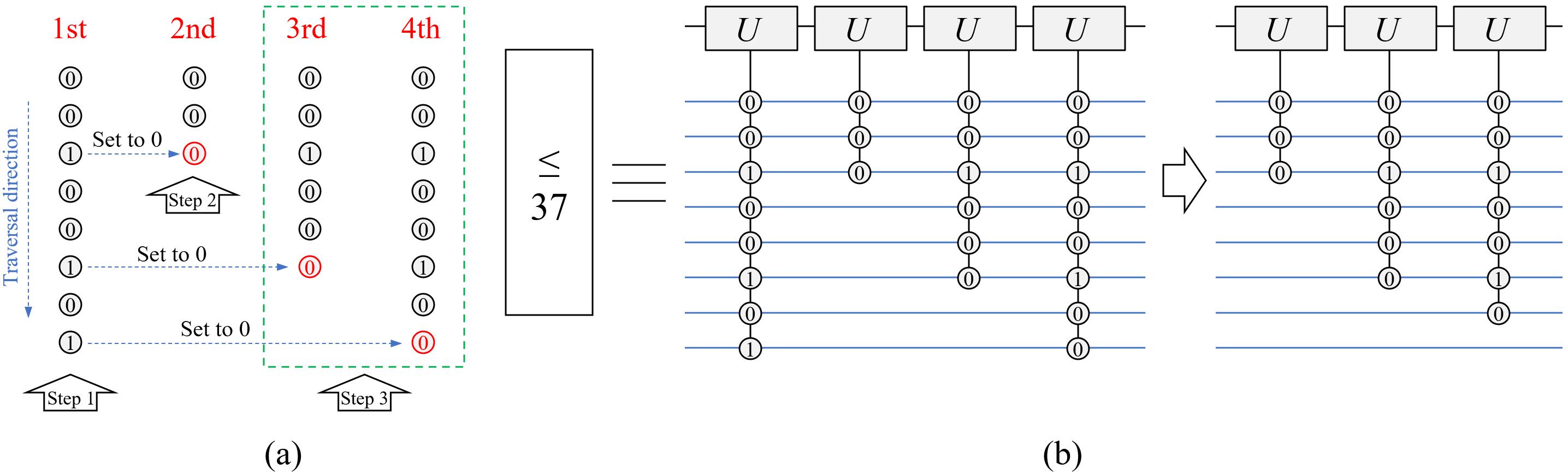}
  \caption{(a) Detailed steps for identifying the pixels in an image with a lightness value less than or equal to 37. (b) A quantum circuit diagram used to transform specified color content.}
  \label{fig11}
\end{figure}

Images can be partitioned and specific pixels can be transformed using control conditions applied to a lightness or position qubit sequence. For example, the pixels in an image with a lightness value less than or equal to $\xi$ = 37 can be identified using the method shown in Figure \ref{fig11}(a), which consists of three steps.

\textbf{Step 1}: The 8-qubit binary string 00100101 (37) forms the first series of control conditions for locating pixels with a lightness value equal to 37.

\textbf{Step 2}: The binary string defined in Step 1 is traversed from high to low until the first instance of 1 is located. It is then set to 0 and serves as the second series of control conditions.

\textbf{Step 3}: This process, traversing the string to identify the \emph{n}-th instance of 1 (which serves as the (\emph{n}+1)-th control condition), is repeated until all pixels with a lightness value less than 37 are identified.

These steps, combined with controlled-rotation gates, can be used to complete the desired operations within a specified lightness range, as shown in Figure \ref{fig11}(b). It is evident that the quantum circuit diagram can be further simplified when the lowest qubit in the string is 1.

\begin{figure}[h]
  \centering
  \includegraphics[width=15cm]{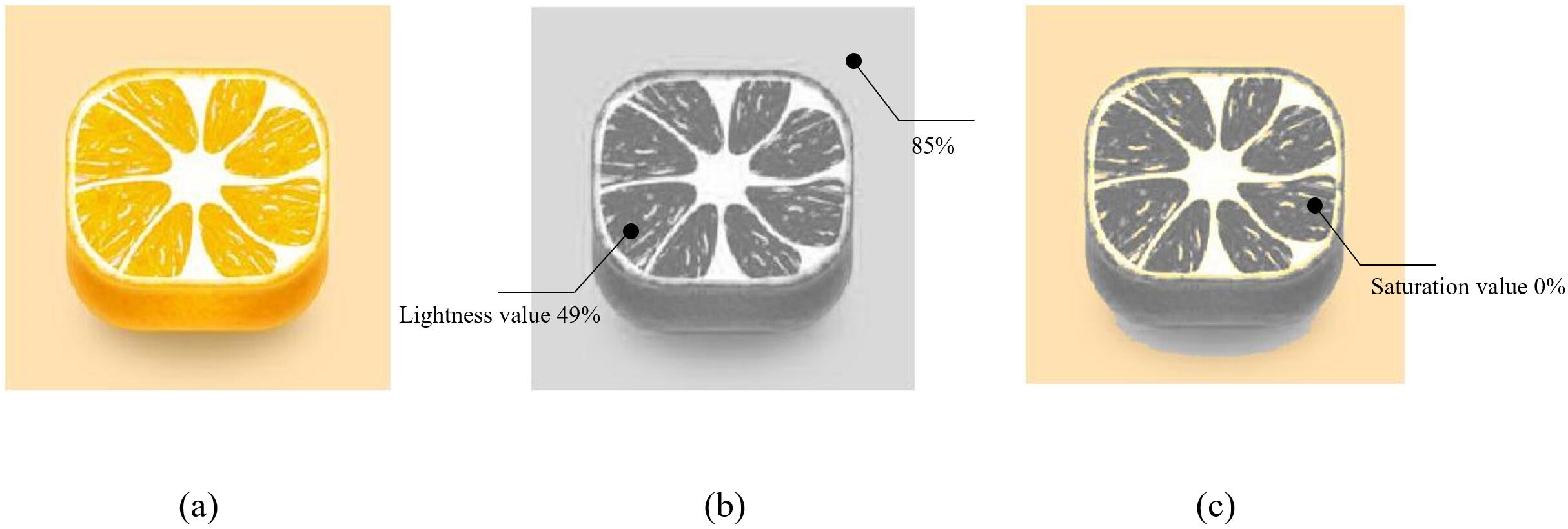}
  \caption{(a) A QHSL image of an orange. (b) Its lightness component. (c) The effects of varying saturation.}
  \label{fig12}
\end{figure}

Figure \ref{fig12}(a) shows a QHSL image with a uniform background and a complex orange pattern in the foreground. Figure \ref{fig12}(b) is the component image from the lightness channel, with a value of 85\% in the background and 49\% in the orange pattern region. An intermediate lightness value of 75\%, mapped to an 8-qubit string of 11000000 using the averaging technique discussed in Section \ref{sec2-2}, was selected to distinguish the foreground from the background and reduce circuit complexity. The orange pattern information was extracted from the image using the three steps discussed previously.

Saturation of the orange pattern can be modified without changing the image background using control conditions and rotation gates $R_Y(\Delta\theta)$. This effect is shown in Figure \ref{fig12}(c) for $\Delta\theta=-\pi/3$, in which the orange pattern saturation reaches 0 and loses its color.

\begin{figure}[h]
  \centering
  \includegraphics[width=16.5cm]{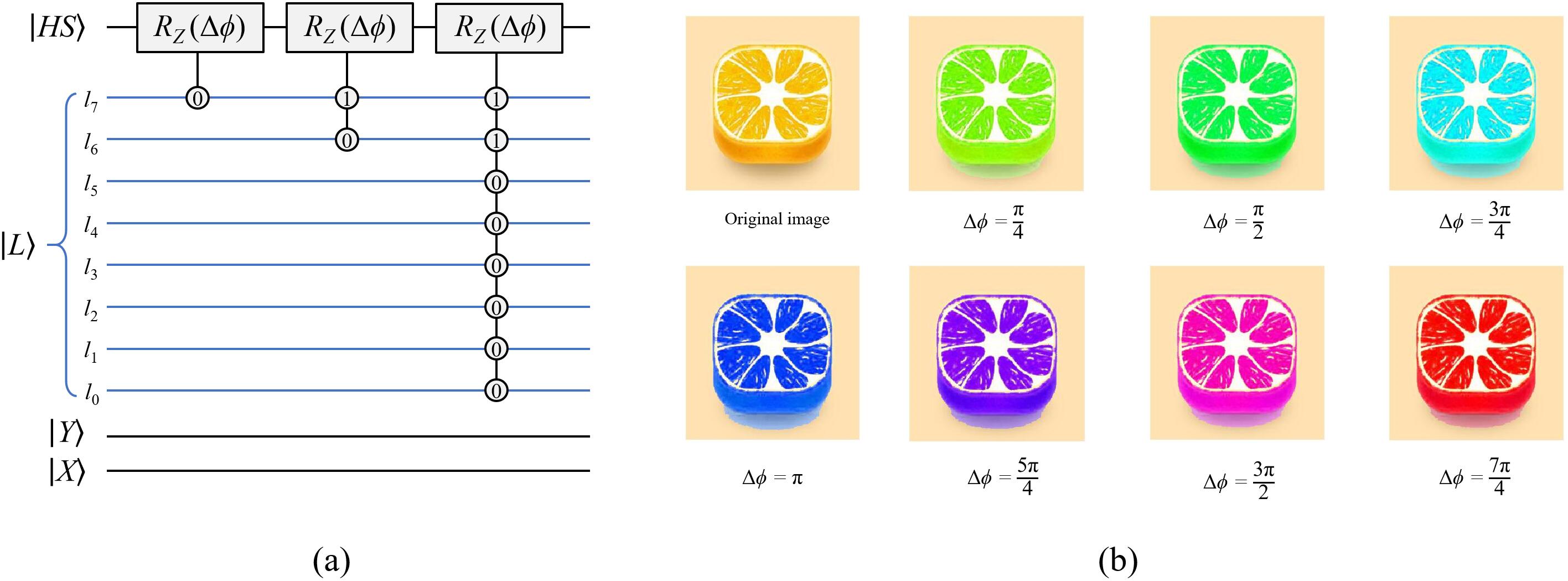}
  \caption{Changes in local hue information for varying lightness values.}
  \label{fig13}
\end{figure}

Hue information can also be modified without affecting the background, using control conditions and rotation matrices $R_Z(\Delta\phi)$. The specific quantum circuit diagram for this process is shown in Figure \ref{fig13}(a). The results of varying $\Delta\phi$ are included in Figure \ref{fig13}(b), which shows several different orange color patterns.

\begin{figure}[h]
  \centering
  \includegraphics[width=15cm]{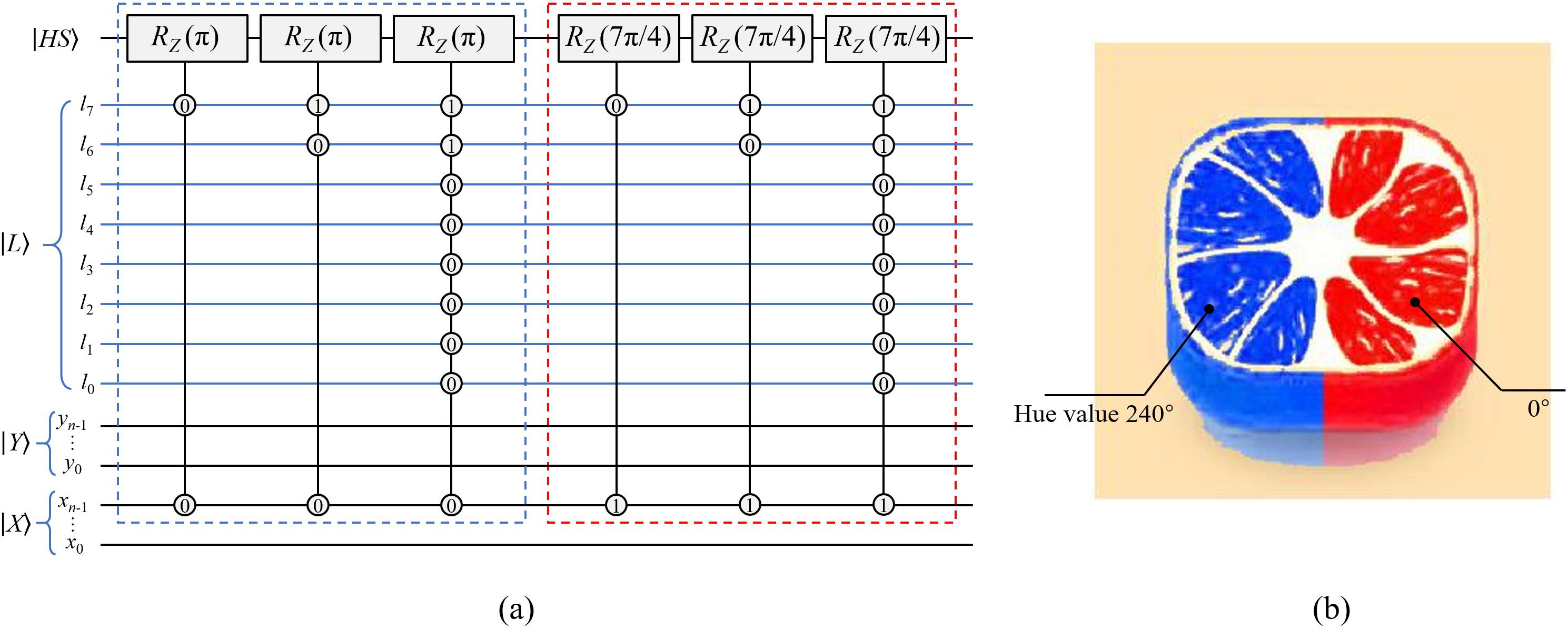}
  \caption{Changes in local hue values using location and lightness information.}
  \label{fig14}
\end{figure}

More complex effects occur when the image is segmented using position and lightness information simultaneously. Figure \ref{fig14}(a) shows the quantum circuit diagram for this combined process, in which the orange pattern can be evenly divided into two parts along the Y-axis by applying constraints to location information. Additional operations, discussed previously, were used to produce the two different colors shown in Figure \ref{fig14}(b).

Several previous studies have modeled changes in circuit complexity as control conditions were added to restricted operations. For example, Le et al. investigated the relationship between the number of control conditions and the size of the affected area in a quantum image, concluding that increasing controls reduced the area \cite{le2011strategies}. In addition, specifying the area in which a transformation will be applied increases the complexity of a transformation by increasing the depth and number of gates in the corresponding circuit.

(2) Local operations with quantum comparators

In addition to local operations using the simple control conditions discussed earlier, pixels can also be constrained with quantum comparator modules used in arithmetic computing \cite{oliveira2007quantum}. Comparators are used to compare two input quantum states, outputting the state with the larger magnitude. This device can be used to accurately and conveniently identify the pixels in a QHSL image that satisfy lightness and position requirements.

\begin{figure}[h]
  \centering
  \includegraphics[width=15cm]{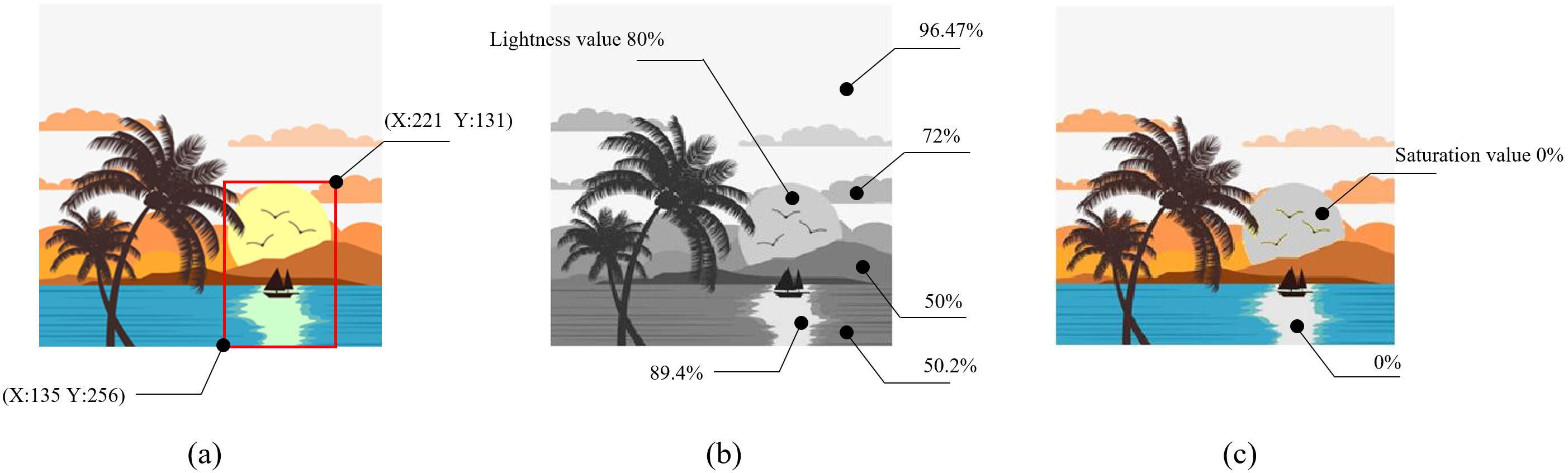}
  \caption{(a) A QHSL image of a natural scene (illustration). (b) The corresponding lightness component. (c) The effects of varying saturation for the sun and its reflection on the water.}
  \label{fig15}
\end{figure}

Figure \ref{fig15}(a) shows a $256\times256$ QHSL image in which the color of the sun and its reflection on the water were modified. The lightness component (i.e., grayscale version) is shown in Figure \ref{fig15}(b). The lightness of the sun and its reflection are in the interval between 76\% ($0.76\times 255\approx194$ shades of gray) and 94.5\% ($0.945\times 255\approx241$ shades of gray). Position information (the coordinate range) was collected for these structures as shown in Figure \ref{fig15}(a)). The saturation was then modified by constraining pixel values using a quantum comparator module. The effects of applying the $R_Y(\Delta\theta=-\pi/3)$ operator to the specified saturation information are evident in Figure \ref{fig15}(c).

\begin{figure}[h]
  \centering
  \includegraphics[width=16cm]{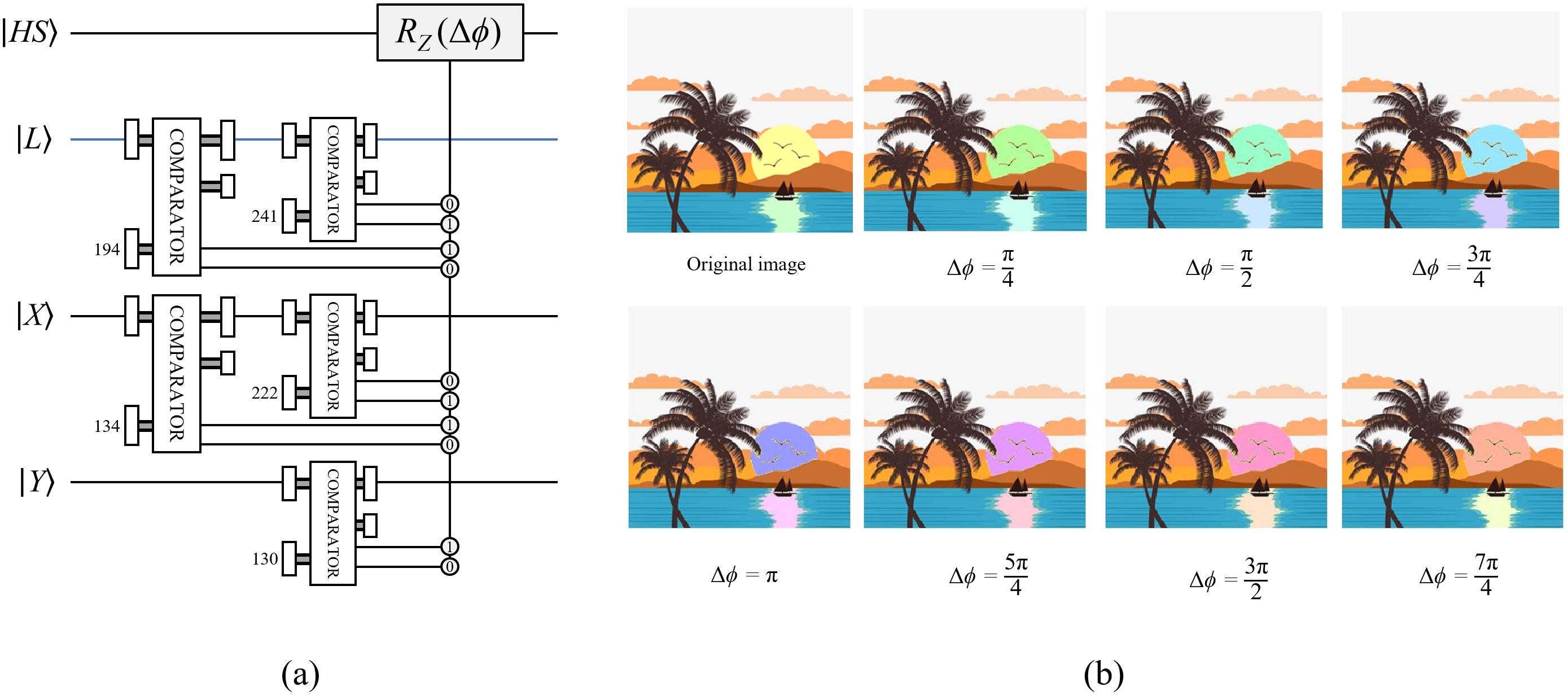}
  \caption{Changes in local hue information using quantum comparators on position information and lightness values.}
  \label{fig16}
\end{figure}

This same approach can be used to alter the hue of specific regions in a QHSL image. The required rotation matrix $R_Z(\Delta\phi)$ and the corresponding quantum circuit are shown in Figure \ref{fig16}(a). The effects of varying $\Delta\phi$ are shown in Figure \ref{fig16}(b), where the sun is displayed in different colors.

\section{Pseudocolor processing with QHSL images}\label{sec4}
Pseudocolor processing can be used to convert an image from grayscale to color or from a monochromatic representation to a specific color distribution \cite{gonzalez2009digital}. This process can aid in distinguishing fine visual details during image enhancement. As discussed previously, the lightness component of a QHSL image is similar to a traditional grayscale image for a saturation value of 0. The primary difference is that grayscale pixel values are typically integers ranging from 0 (black) to 255 (white). However, lightness values vary continuously from $0-100\%$. As a result, if both the grayscale and lightness values are stored in a quantum sequence composed of 8 qubits, they are equivalent in bitwise inversion.

Pseudocolor processing using NEQR images has previously been implemented using quantum color maps \cite{jiang2015quantum}. However, this approach requires the gray pixel depths in the original image to be evenly divided into several sub-intervals, which causes the algorithm to fail in other cases. In this study, the proposed pseudocolor processing can be achieved using uneven divisions of gray depths, which makes images more vivid and further highlights fine details.

\begin{figure}[h]
  \centering
  \includegraphics[width=14cm]{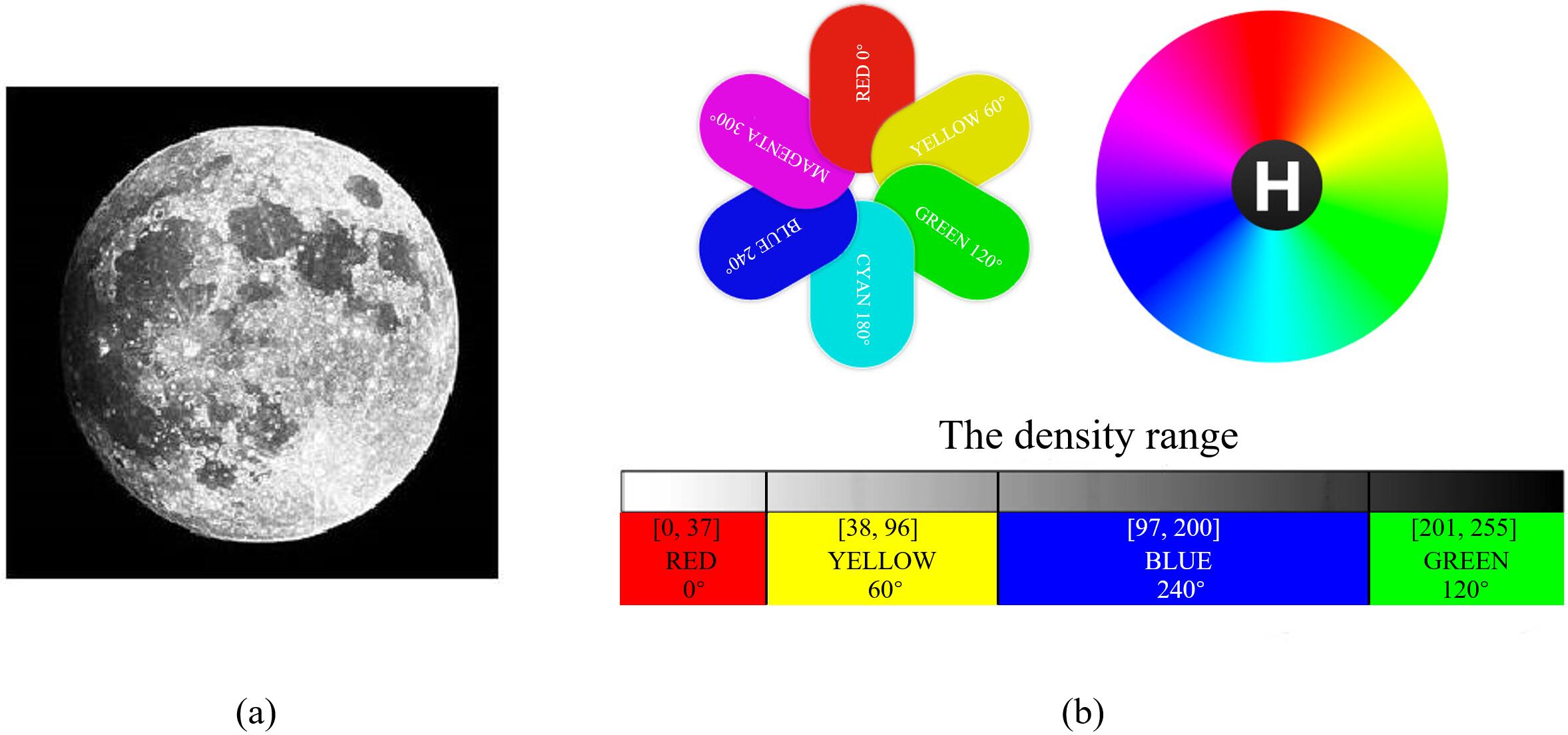}
  \caption{(a) The initial lunar image. (b) The angular positions of 4 colors in the hue circle and their corresponding density range.}
  \label{fig17}
\end{figure}

The image of the moon in Figure \ref{fig17}(a) provides an example of pseudocolor processing. Figure \ref{fig17}(b) includes a lookup table used to divide the density range [0, 255] into the 4 uneven sub-intervals [0, 37], [38, 96], [97, 200], and [201, 255]. These ranges were assigned to red, yellow, blue, and green, respectively. The angular positions of the 4 colors in the hue circle are also shown in Figure \ref{fig17}(b).

The quantum pseudocolor method proposed in this study can be divided into the following three steps.

\textbf{Step 1}: As discussed in Section \ref{sec3-1}, grayscale information in Figure \ref{fig17}(a) was encoded using 8 qubits and compiled into the lightness component of a QHSL image.

\textbf{Step 2}: Four rotation operations $R_Z(\Delta\phi_i)$ $(i=0,1,2,3)$ must be applied to the hue components in order to add the four specified colors to different sub-intervals of the density range shown in Figure \ref{fig17}(b). This process involves rotating all pixels less than or equal to 37 using $R_Z(\Delta\phi_0)$ and rotating all pixels less than or equal to 96 (including all pixels less than or equal to 37) using $R_Z(\Delta\phi_1)$. The terms $\Delta\phi_i$ $(i=0,1,2,3)$ must then be calculated in reverse order. All pixels, beginning with the largest sub-interval [201, 255], were converted to green. This corresponds to an angle of $120\degree$ or $\Delta\phi_3=2\pi/3$. Proceeding to the lower sub-interval, all pixels with grayscale values in the range [97, 200] were converted to blue. This corresponds to an angle of $240\degree$ or $4\pi/3$. Rotation operations in higher sub-intervals affect pixels in lower sub-intervals. As such, $\Delta\phi_3$ should be added to calculate $\Delta\phi_2$ (i.e., $\Delta\phi_2+\Delta\phi_3=4\pi/3$), $\Delta\phi_1$, and $\Delta\phi_0$. This process for calculating the four rotation angles $\Delta\phi_i$ $(i=0,1,2,3)$ can be described as follows:

\begin{equation}\label{eq25}
 \begin{aligned}
&\left\{
\begin{aligned}
 &\Delta\phi_3=\frac{2\pi}{3}\\
 &\Delta\phi_2+\Delta\phi_3=\frac{4\pi}{3}\\
 &\Delta\phi_1+\Delta\phi_2+\Delta\phi_3=\frac{\pi}{3}\\
 &\Delta\phi_0+\Delta\phi_1+\Delta\phi_2+\Delta\phi_3=0\\
 \end{aligned}
\right.\Rightarrow
&\left\{
\begin{aligned}
 &\Delta\phi_0=-\frac{\pi}{3}\\
 &\Delta\phi_1=-\pi\\
 &\Delta\phi_2=\frac{2\pi}{3}\\
 &\Delta\phi_3=\frac{2\pi}{3}\\
 \end{aligned}
\right..\\
\end{aligned}
\end{equation}

\begin{figure}[h]
  \centering
  \includegraphics[width=16cm]{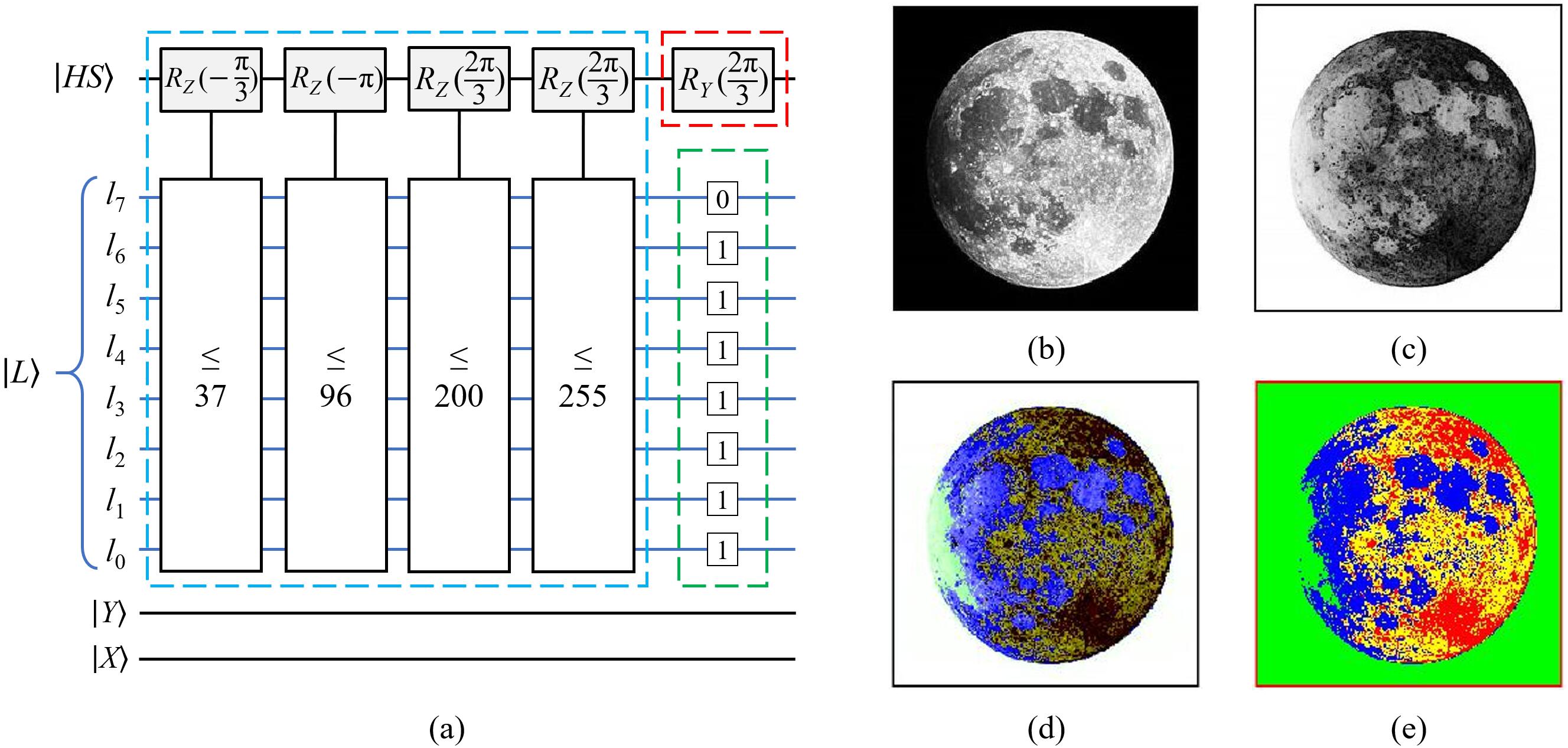}
  \caption{(a) The quantum circuit diagram for the process described in Steps 1-3. Also shown are (b) the original lunar image, (c) the effect of Step 2, (d) and (e) the effect of Step 3.}
  \label{fig18}
\end{figure}

The quantum circuit diagram for this process is shown in the blue dotted box of Figure \ref{fig18}(a). The completion of this step satisfies all hue storage requirements, but the saturation component is still 0\% so the image remains a grayscale representation without obvious colors, as shown in Figure \ref{fig18}(c).

\textbf{Step 3}: The four colors selected in the pseudocolor processing were all pure colors requiring a saturation of 100\%, set using the $R_Y(\Delta\theta=2\pi/3)$ operation, as seen in the red dotted box of Figure \ref{fig18}(a). The QHSL image produced after changing the saturation is shown in Figure \ref{fig18}(d). The lightness value must simultaneously be set to 50\% using quantum set 0 and 1 gates. A lightness value of 50\% is represented by the 8-qubit binary number 01111111 (127) because 127/255 = 0.498 (rounded to 50\%), as shown in the green dotted box of Figure \ref{fig18}(a). The QHSL image produced after Step 3 is shown in Figure \ref{fig18}(e).

\begin{figure}[h]
  \centering
  \includegraphics[width=15cm]{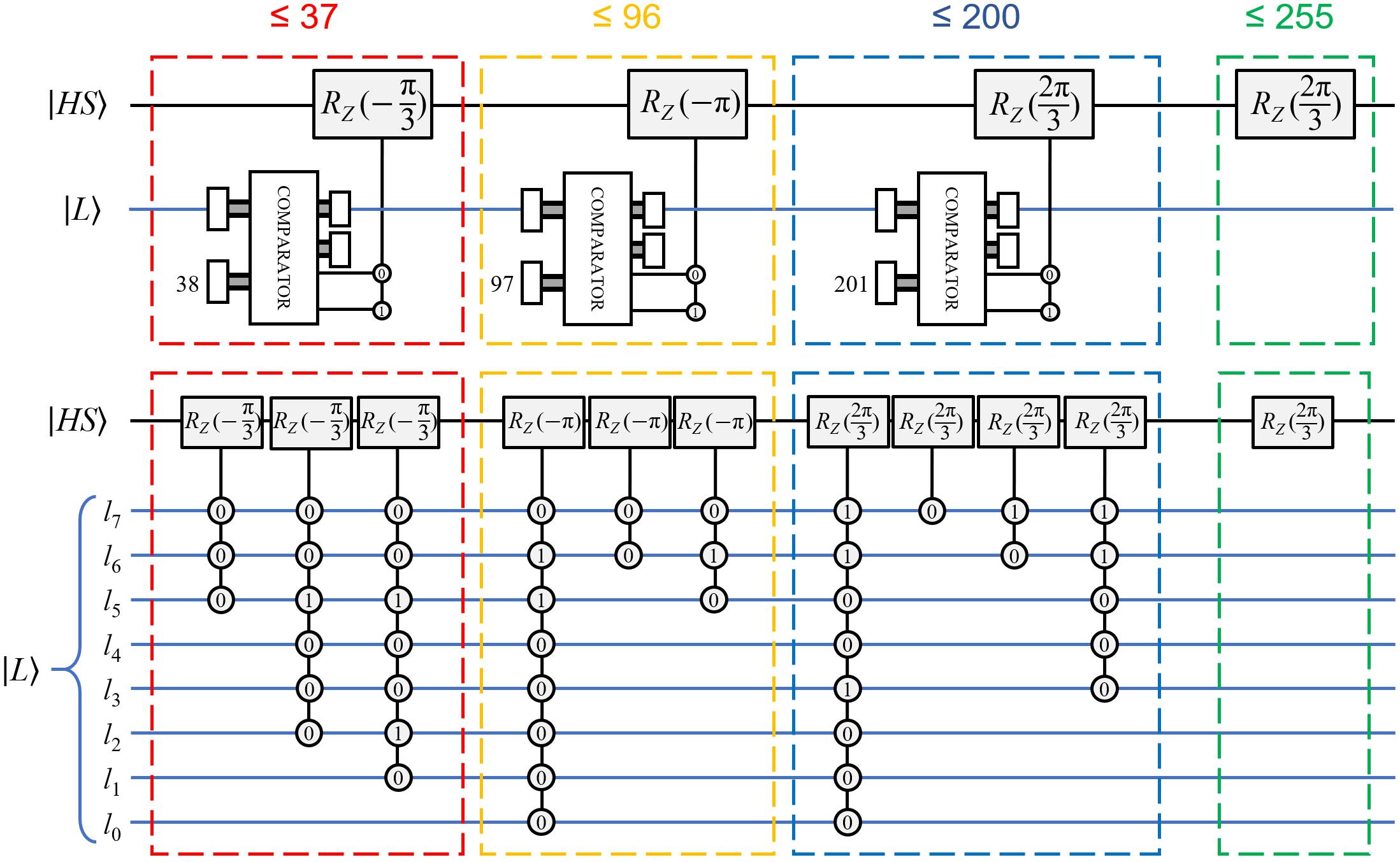}
  \caption{Two possible quantum circuits for the pseudocolor processing described in Step 2.}
  \label{fig19}
\end{figure}

As discussed previously, there are two options for identifying grayscale pixels less than or equal to a certain value, as shown in the blue dotted box of Figure \ref{fig18}(a). The quantum circuit diagram described in Step 2, composed of quantum comparators, is shown in the top half of Figure \ref{fig19}. The circuit developed using control conditions is shown in the lower half of the figure. It is concluded that the upper circuit offers higher accuracy, while the lower circuit exhibits lower computational complexity.

\section{Conclusion}\label{sec5}
To embody information in the features of physical support will decide the power level of a certain computational paradigm. Quantum computation has emerged as a powerful tool for efficient and secure information processing. The same with the effort to process any type of information, the formalized encoding is a must. This study proposed a novel quantum hue, saturation, and lightness (QHSL) color model for the representation and storage of images. This work represents the first attempt to develop an HSL color model in the field of quantum computing, thus providing a new impetus to investigate the production and manipulation of quantum images in terms of triple perceptually relevant components. A pseudocolor technique, based on a density stratification method, was also presented. This approach is advantageous because, unlike conventional algorithms, it is capable of pseudocolor processing with uneven gray depth divisions for highlighting fine image details.

In future work, the results from this study can be extended towards the following directions. Firstly, although the selection of 8 qubits for encoding lightness information in the QHSL image allows for more operations in image processing, the trade-off between qubit storage and additional applications should be studied in greater detail. Secondly, the possibility of using quantum-based image processing algorithms, such as quantum Fourier transforms \cite{yang2014quantumcryp} and wavelet transforms \cite{li2020aquantummechanics}, for advanced image manipulation should also be explored further. Finally, basic quantum image manipulation could serve as a foundation for more sophisticated algorithms used in applications such as watermarking and steganography \cite{el-latif2019efficient}, in which QHSL images could be a valuable new tool for information processing.


\end{spacing}
\end{document}